\documentclass[preprint,sort&compress]{elsarticle}
\usepackage{amssymb}
\usepackage{amsthm}
\usepackage{amsmath}
\usepackage{bm}
\usepackage{multirow}
\usepackage{hepunits}
\usepackage[svgnames]{xcolor}
\usepackage{hyperref}
\usepackage{multirow}
\journal{Physics Letters B}
\begin{document}

\begin{frontmatter}

\title{Next-to-leading order corrections for $gg \to ZH$ with top quark mass dependence}

\author[1]{Guoxing Wang}
\ead{wangguoxing2015@pku.edu.cn}
\author[2]{Xiaofeng Xu}
\ead{pkuxxf@gmail.com}
\author[3]{Yongqi Xu}
\ead{xuyongqi@pku.edu.cn}
\author[1]{Li Lin Yang}
\ead{yanglilin@zju.edu.cn}

\address[1]{Zhejiang Institute of Modern Physics, Department of Physics, Zhejiang University, Hangzhou 310058, China}
\address[2]{Institut f\"ur Theoretische Physik, Universit\"at Bern, Sidlerstrasse 5, CH-3012 Bern, Switzerland}
\address[3]{School of Physics and State Key Laboratory of Nuclear Physics and Technology, Peking University, Beijing 100871, China}

\begin{abstract}
In this work, we present for the first time a calculation of the complete next-to-leading order corrections to the $gg \to ZH$ process. We use the method of small mass expansion to tackle the most challenging two-loop virtual amplitude, in which the top quark mass dependence is retained throughout the calculations. We show that our method provides reliable numeric results in all kinematic regions, and present phenomenological predictions for the total and differential cross sections at the Large Hadron Collider and its future upgrades. Our results are necessary ingredients towards reducing the theoretical uncertainties of the $pp \to ZH$ cross sections down to the percent-level, and provide important theoretical inputs for future precision experimental collider programs.
\end{abstract}

\begin{keyword}
Higgs boson \sep Z boson \sep QCD

\end{keyword}

\end{frontmatter}

%% \linenumbers

%% main text
\section{Introduction}

One of the top priorities of the Large Hadron Collider (LHC) and the High Luminosity LHC (HL-LHC) is to accurately measure the properties of the Higgs boson including its various couplings. In particular, the Yukawa couplings of the Higgs boson to fermions are key to understand the origin of fermion masses as well as to verify the family universality of different fermion generations. At the LHC and the HL-LHC, the main production channel $gg \to H$ followed by hadronic decays of the Higgs boson is overwhelmed by large hadronic activities arising from strong interactions. Therefore, to measure the Yukawa couplings of quarks lighter than the top quark, the $pp \to ZH$ production channel is the most promising one, where the leptonic decay of the $Z$ boson can be efficiently detected and leads to a significant reduction of the hadronic background \cite{ATLAS:2018kot, CMS:2018nsn}. Besides, the $pp \to ZH$ process also provides an independent handle of the $HZZ$ coupling with respect to the $H \to ZZ^*$ decay process\cite{LHCHiggsCrossSectionWorkingGroup:2013rie,CMS:2018uag}. Therefore, it is of crucial importance to have precise knowledge of its differential cross sections both from the theoretical and the experimental points of view.

At the partonic level, the $pp \to ZH$ process proceeds through the subprocess $q\bar{q} \to Z^* \to ZH$ at the leading order (LO) in standard model (SM) couplings. Since the quantum effects due to strong interactions are often quite large for such processes, it is necessary to calculate higher order corrections in the strong coupling $\alpha_s$ within perturbative quantum chromodynamics (QCD), in order to improve the accuracy for the theoretical predictions. The next-to-leading order (NLO) and next-to-next-to-leading order (NNLO) QCD corrections were calculated in \cite{Han:1991ia, Brein:2003wg,Brein:2011vx} and are implemented in the program package \texttt{vh@nnlo}~\cite{Brein:2012ne, Harlander:2018yio}. The NLO electroweak (EW) corrections are given by \cite{Ciccolini:2003jy, Denner:2011id}. Combined QCD+EW corrections at NLO are also available \cite{Denner:2014cla, Granata:2017iod, Obul:2018psx}. Despite of these progresses, the theoretical precision is still not enough to match the potential of future LHC and HL-LHC runs~\cite{Cepeda:2019klc}.

There is one special class of higher order corrections coming from the partonic subprocess $gg \to ZH$ mediated by a closed top quark loop. Such contributions enter at order $\alpha_s^2$ in perturbation theory. However, this formal suppression by the coupling constant is compensated partly by the high luminosity of gluons at colliders such as the LHC and the HL-LHC, and partly by the large Yukawa coupling of the top quark. Moreover, the differential cross sections for this subprocess receive additional enhancement by the $2m_t$ threshold effects in the boosted regime, e.g., when the Higgs transverse momentum $p_{T} \geq 150~\mathrm{GeV}$ \cite{Harlander:2013mla, Englert:2013vua}.
For the total cross section at the 13~TeV LHC, the one-loop $gg \to ZH$ contribution is roughly 50\% in size compared to the NLO corrections for the $q\bar{q} \to ZH$ channel, and is much larger than the NNLO corrections for the $q\bar{q}$ channel.
Therefore, it is reasonable to expect that the formally $\alpha_s^3$ contributions in the $gg \to ZH$ category are not small compared to the formally $\alpha_s^2$ terms in the $q\bar{q}$ channel. In fact, the estimate in the heavy top limit \cite{Altenkamp:2012sx, Hasselhuhn:2016rqt} suggests that, the $gg \to ZH$ contribution at order $\alpha_s^3$ is similar in size to that at order $\alpha_s^2$. In the case of Higgs boson pair production, it was realized that the heavy top limit is not quite reliable, and the finite top mass effects are significant \cite{Hasselhuhn:2016rqt}. As a result, the unknown size of the exact $gg \to ZH$ contribution at order $\alpha_s^3$ has become a major uncertainty of the theoretical predictions for the $pp \to ZH$ cross sections~\cite{Cepeda:2019klc}.

Last but not least, the loop-induced $gg\to ZH$ subprocess is interesting on its own right. It is sensitive to possible new heavy particles running in the loop. From the low energy point of view, it has a different dependence on the $HZZ$ anomalous coupling than the tree-mediated $q\bar{q}$ channel. It is also sensitive to other anomalous couplings such as $Ht\bar{t}$ and $Zt\bar{t}$. This channel therefore provides unique information on the possible new physics beyond the SM.

The above considerations make it highly interesting to calculate the $gg \to ZH$ subprocess to the next order in perturbation theory. With a slight abuse of notation (as adopted in the literature), we will refer to the one-loop amplitude-squared contributions as the ``LO'' cross sections for this process. These have been computed in \cite{Dicus:1988yh, Kniehl:1990iva}. The ``NLO'' corrections to this process then consist of two parts: the interference between the two-loop and one-loop amplitudes (the virtual contributions), and the squared-amplitude with one loop and one extra parton radiation (the real contributions). Both contributions are infrared (IR) divergent, while their sum leads to a finite prediction. The bottleneck in such a calculation lies in the two-loop virtual amplitude, which involves multiple physical scales including three masses ($m_Z$, $m_H$ and $m_t$) and two Mandelstam variables.

The presence of multiple physical scales makes the computation of the two-loop amplitude rather challenging. Very recently, a purely numeric study based on sector decomposition \cite{Hepp:1966eg,Roth:1996pd,Binoth:2000ps,Heinrich:2008si} has appeared \cite{Chen:2020gae}. Approximations in certain kinematic regions have also been worked out, including the high energy expansion \cite{Davies:2020drs} and the small transverse momentum expansion \cite{Alasfar:2021ppe}. In \cite{Xu:2018eos}, some of the authors of this work proposed a novel approach to calculate loop integrals with a heavy top quark loop and lighter external particles such as the Higgs boson and weak gauge bosons, which is valid in all phenomenologically relevant kinematic regions.
In this work, we apply this method to calculate the NLO virtual corrections to the process $gg\to ZH$. We show that our method gives accurate numeric results for the two-loop amplitudes in the entire physical phase-space. Combining with the real corrections, we for the first time provide the complete NLO predictions for the total and differential cross sections of the $gg \to ZH$ channel. We also give the total cross sections for the full $pp \to ZH$ process by adding the $q\bar{q}$ contributions.

\section{Setup and analytic calculations}

We consider the partonic process $g^{\alpha}_{a}(p_1)+g^{\beta}_{b}(p_2) \to Z^{\mu}(p_3)+H(p_4)$, where $a$ and $b$ are color indices, while $\alpha$, $\beta$ and $\mu$ are Lorentz indices. The amplitude can be written as
\begin{align}
\mathcal{M}_{ab}^{\alpha\beta\mu} &= \frac{G_F}{\sqrt{2}} \frac{\alpha_s}{2\pi} \, \hat{s} \, m_Z\, \delta_{ab} \, \sum_{i=1}^{7} A_i^{\alpha\beta\mu} F_i \, ,
\label{eq:amp}
\end{align}
where the tensor structures $A_i^{\alpha\beta\mu}$ are given in \cite{Kniehl:1990iva}, whose coefficients $F_i$ are functions of the masses $m_t$, $m_Z$, $m_H$ and the Mandelstam variables $\hat{s}=(p_1+p_2)^2$, $\hat{t}_1=(p_1-p_3)^2-m_Z^2$, $\hat{u}_1=(p_1-p_4)^2-m_H^2$. These variables satisfy $\hat{s} + \hat{t}_1 + \hat{u}_1 = 0$. For later convenience, we also define $\hat{t}=\hat{t}_1+m_Z^2$ and $\hat{u}=\hat{u}_1+m_H^2$.

To calculate the squared-amplitude, we multiply Eq.~\eqref{eq:amp} by its complex conjugate, and sum over the polarization states of the gluon and the $Z$ boson. For simplicity, we define the coefficients $\hat{C}_i \equiv \sum_{j=1}^{7} M_{ij} F_j$, where the matrix elements $M_{ij}$ are given by
\begin{equation}
M_{ij} = - \eta_{\alpha\rho} \, \eta_{\beta\sigma}\left(\eta_{\mu\nu} - \frac{p_{3\mu}p_{3\nu}}{m_Z^2}\right) A_i^{\alpha\beta\mu} A_j^{\rho\sigma\nu} \, .
\label{eq:Mij}
\end{equation}

The coefficients $\hat{C}_i$ can be expanded in terms of the strong coupling constant $\alpha_s$:
$\hat{C}_i = \hat{C}_i^{(0)} + (\alpha_s/4\pi) \, \hat{C}_i^{(1)} + \cdots$,
where $\hat{C}_i^{(0)}$ are the LO contributions, and $\hat{C}_i^{(1)}$ are the NLO virtual corrections, which involve complicated two-loop Feynman integrals with massive external legs. In the following, we describe the calculation of $\hat{C}_i^{(1)}$ using the method of small mass expansion.

We generate the relevant two-loop virtual diagrams and the corresponding amplitudes using \texttt{FeynArts} \cite{Hahn:2000kx}. The resulting expressions are manipulated with \texttt{FeynCalc} \cite{Shtabovenko:2016sxi} and \texttt{FORM} \cite{Vermaseren:2000nd}. The two-loop diagrams can be classified into several categories: diagrams consisting of two one-loop sub-diagrams, two-loop triangle diagrams (involving both top quark and bottom quark loops attached to an offshell $Z$ boson), and two-loop box diagrams. While the first two categories can be easily calculated \cite{Hasselhuhn:2016rqt,Davies:2020drs,Spira:1995rr,Graudenz:1992pv,Gehrmann:2005pd}, the two-loop box diagrams are rather challenging. We denote the corresponding contributions to the $\hat{C}_i^{(1)}$ coefficients as $\hat{C}^{(1)}_{i,\text{box}}$, and calculate them using the small mass expansion. Namely, we expand $\hat{C}^{(1)}_{i,\text{box}}$ as a power series in $m_Z^2$ and $m_H^2$:
\begin{align}
\label{eq:expansionm}
&\hat{C}_{i,\text{box}}^{(1)}(\hat{s},\hat{t}_1,m_t^2,m_Z^2,m_H^2) =  \frac{1}{m_Z^2} \, \sum_{n=0}^{\infty} \sum_{l=0}^{n} \, \frac{(m_Z^2)^l}{l!} \nonumber
\\
&\times \frac{(m_H^2)^{n-l}}{(n-l)!} \, \left[ \frac{\partial^{n} \left(m_Z^2 \hat{C}_{i,\text{box}}^{(1)}\right)}{\partial (m_Z^2)^l \, \partial (m_H^2)^{n-l}} \right]_{m_Z^2,m_H^2 \to 0} \, .
\end{align}
It should be noted that in contrast to the case of Higgs boson pair production \cite{Xu:2018eos, Wang:2020nnr}, there is an extra factor of $1/m_Z^2$ due to the polarization sum of the $Z$ boson, as is evident in Eq.~\eqref{eq:Mij}. For the power counting we regard $m_Z^2 \sim m_H^2$, and denote the $n$th term ($n \geq 0$) in the expansion as order $m^{2(n-1)}$. We will also use the notation $\mathcal{O}(m^{2(n-1)})$ to denote the partial sum of the series up to the $n$th term. In practice, we have performed the expansion up to $n = 3$, corresponding to $\mathcal{O}(m^4)$.

Acting on the amplitudes, the partial derivatives $\partial_{m_Z^2}$ and $\partial_{m_H^2}$ can be written as partial derivatives with respect to the external momenta:
\begin{align}
\partial_{m_Z^2} &= \frac{\hat{s} (p_T^2 - \hat{u}) p_1^\mu + \hat{t}_1 \hat{u}_1 p_2^\mu + \hat{s}\hat{u}_1 p_3^\mu}{2 \hat{s}^2 p_T^2} \, \partial_{p_3^\mu} \nonumber
\\
\partial_{m_H^2} &= \frac{\hat{s} (m_Z^2-p_T^2) p_1^\mu + \hat{t}^2_1 p_2^\mu + \hat{s}\hat{t}_1 p_3^\mu}{2 \hat{s}^2 p_T^2} \, \partial_{p_3^\mu}
\, ,
\label{eq:derivative}
\end{align}
where $p_T^2=\hat{t}\hat{u}/\hat{s}-m_Z^2m_H^2/\hat{s}$. Setting $m_Z^2,m_H^2 \to 0$ after taking the derivatives, the coefficients $\hat{C}^{(1)}_{i,\text{box}}$ can be written as linear combinations of scalar loop integrals with massless external legs. These integrals have already been computed in \cite{Caron-Huot:2014lda, Becchetti:2017abb, Xu:2018eos, Wang:2020nnr}.

The loop integrals contain both ultraviolet (UV) and infrared (IR) divergences which are regularized in dimensional regularization. The UV divergences are removed via renormalization. We adopt the on-shell scheme for $m_t$ and the five-flavor $\overline{\rm{MS}}$ scheme for $\alpha_s$. In addition, to handle $\gamma_5$ in the amplitudes, we apply the Larin scheme~\cite{Larin:1993tq} which requires a finite renormalization constant $Z_5 = 1 - C_F\alpha_s/\pi + \mathcal{O}(\alpha_s^2)$.

The IR divergences will be eventually cancelled by the real corrections and the renormalization of the parton distribution functions (PDFs). We apply the Catani-Seymour dipole subtraction method \cite{Catani:1996vz} to subtract the IR divergences from the virtual corrections. After that, we define the UV- and IR-finite coefficients $\hat{C}^{(1)}_{i,\rm{fin}} =  \hat{C}^{(1)}_{i,\rm{ren}} -  \bm{I} \otimes\hat{C}^{(0)}_{i}$, where $\hat{C}^{(1)}_{i,\rm{ren}}$ are the renormalized coefficients and the IR subtraction operator $\bm{I}$ is defined as
\begin{equation}
\bm{I}(\epsilon) =  - 2 \, \frac{(4\pi)^{\epsilon}}{\Gamma(1-\epsilon)} \left(\frac{\mu_r^2}{-\hat{s}}\right)^{\epsilon} \left[ \frac{C_A}{\epsilon^2} + \frac{\beta_0}{2\epsilon} + \tilde{K}_g \right] ,
\label{eq:Ioperator}
\end{equation}
where $\beta_0=11C_A/3-4T_FN_l/3$ with $T_F=1/2$ and $N_l=5$, and $\tilde{K}_g$ is the finite part and is given by
$\tilde{K}_g = 67C_A/18 - 10T_FN_l/9 + \beta_0/2$.

The contributions of the finite coefficients $\hat{C}^{(1)}_{i,\rm{fin}}$ to the squared-amplitude can be summarized into the so-called finite part of the NLO virtual corrections:
\begin{equation}
\label{eq:vfin}
\mathcal{V}_{\rm fin}= \frac{G_F^2 \, \hat{s}^2 \, m_Z^2}{1024 \pi^2}  \sum_{i,j=1}^{7} 2 \,{\rm Re}\left[\hat{C}^{(0)*}_i \, \left(M^{-1}\right)_{ij} \,  \hat{C}^{(1)}_{j,\rm{fin}} \right]\, .
\end{equation}
In the small mass expansion, the computation of $\mathcal{V}_{\rm fin}$ is rather efficient, which only takes 2 seconds on average for one phase-space point on a workstation with an Intel Xeon W-2155 CPU. Most of the evaluation time is spent on the computation of scalar master integrals. To further accelerate the phase-space integration (which requires to repeatedly evaluate the amplitude), we have generated a very large grid for the master integrals with 63175 points on the two-dimensional plane $(-4m_t^2/\hat{s},-4m_t^2/\hat{t}_1)$. Note that this grid can be reused for different masses (including $m_t$, $m_H$ and $m_Z$), as well as for different couplings ($HZZ$, $Ht\bar{t}$ and $Zt\bar{t}$). Using the grid bypasses the most time-consuming part of the numeric evaluation. As a result, computing the amplitude with the grid only takes $0.003$ seconds on average per phase-space point. 

We now need to add the IR-subtracted real corrections to obtain physical cross sections. We consider squared amplitudes of four partonic subprocesses $gg \to ZH + g$, $gq \to ZH + q$, $g\bar{q} \to ZH + \bar{q}$ and $q\bar{q} \to ZH + g$, and perform the dipole subtraction to remove the IR divergences in the phase-space integration. In the above four subprocesses, we only include the diagrams with a closed quark loop. The IR limit of these diagrams has the same topology as that of the two-loop virtual diagrams. This selection also allows us to compare our results with those in the heavy top limit \cite{Altenkamp:2012sx}. We compute the amplitudes using the package \texttt{Gosam} \cite{Cullen:2011ac,Cullen:2014yla}, and integrate over the phase-space using the Vegas algorithm implemented in the \texttt{Cuba} library \cite{Hahn:2004fe}. To avoid numeric instabilities, we require the transverse momentum of the extra radiation to be larger than $\delta\sqrt{\hat{s}}$. We have varied $\delta$ between $0.05$ and $0.0005$ and find that the result is stable within the integration precision of $0.3\%$.

\section{Numeric results}
In this section, we present our numeric results for the total cross sections and the $M_{ZH}$ distributions, where $M_{ZH} = \sqrt{(p_3 + p_4)^2}$ is the invariant mass of the $Z$ boson and the Higgs boson. For the input parameters, we take $G_F = 1.16637\times10^{-5}$~$\rm{GeV}^{-2}$, $m_Z = 91.1876$~GeV, $m_H = 125$~GeV, $m_t = 172.5$~GeV \cite{ParticleDataGroup:2020ssz}, and use NNPDF3.1 NNLO PDFs \cite{NNPDF:2017mvq} with $\alpha_s(m_Z)=0.118$. We neglect the mass of the bottom quark appearing in the loop. According to \cite{LHCHiggsCrossSectionWorkingGroup:2016ypw}, we set the default values for the renormalization scale $\mu_r$ and the factorization scale $\mu_f$ to be $\mu_{\text{def}} = M_{ZH}$, while the scale uncertainties are estimated by varying the two scales simultaneously up and down by a factor of three.

\begin{table}[t!]
%\centering
\hspace{-4ex}
\footnotesize
\begin{tabular}{r|r|r|rrr}
\multicolumn{1}{c|}{\multirow{2}{*}{$\hat{s}/m_t^2$}} & \multicolumn{1}{c|}{\multirow{2}{*}{$\hat{u}/m_t^2$}} & \multicolumn{4}{c}{$\mathcal{V}'_{\text{fin}}$}
\\
\cline{3-6}
& & \multicolumn{1}{c|}{\texttt{pySecDec}} & \multicolumn{1}{c}{$\mathcal{O}(m^0)$} & \multicolumn{1}{c}{$\mathcal{O}(m^2)$} & \multicolumn{1}{c}{$\mathcal{O}(m^4)$}
\\ \hline
1.707133657190554 &  $-0.441203767016323$ &   35.429092(6) & 35.9823  & 35.5530 & 35.4478 \\
3.876056604162662 &  $-1.616287256345735$ &   4339.045(1)  & 4319.37  & 4336.63 & 4338.73 \\
4.130574250302561 &  $-1.750372271104745$ &   6912.361(3)  & 6870.47  & 6906.92 & 6911.64 \\
4.130574250302561 &  $-2.595461551488002$ &   6981.09(2)   & 6979.28  & 6980.14 & 6980.85 \\
134.5142052093564 &  $-70.34125943305149$ &   $-153.9(4)$  & $-154.543$  & $-154.458$ & $-154.460$ \\
134.5142052093564 &  $-105.1770655376327$ &   527(4)       & 524.585  & 525.958 & 525.965
\end{tabular}
\caption{The finite part of the virtual corrections at six representative phase-space points. The column labelled $\texttt{pySecDec}$ contains results from \cite{Chen:2020gae}, while those labelled $\mathcal{O}(m^n)$ come from the small mass expansion. Note that the numbers correspond to $\mathcal{V}'_{\text{fin}}$ (see the text for explanation).}
\label{tab:amplitudeResults}
\end{table}

\begin{figure}[t!]
\centering
%\hspace{-4ex}
\includegraphics[width=0.6\linewidth]{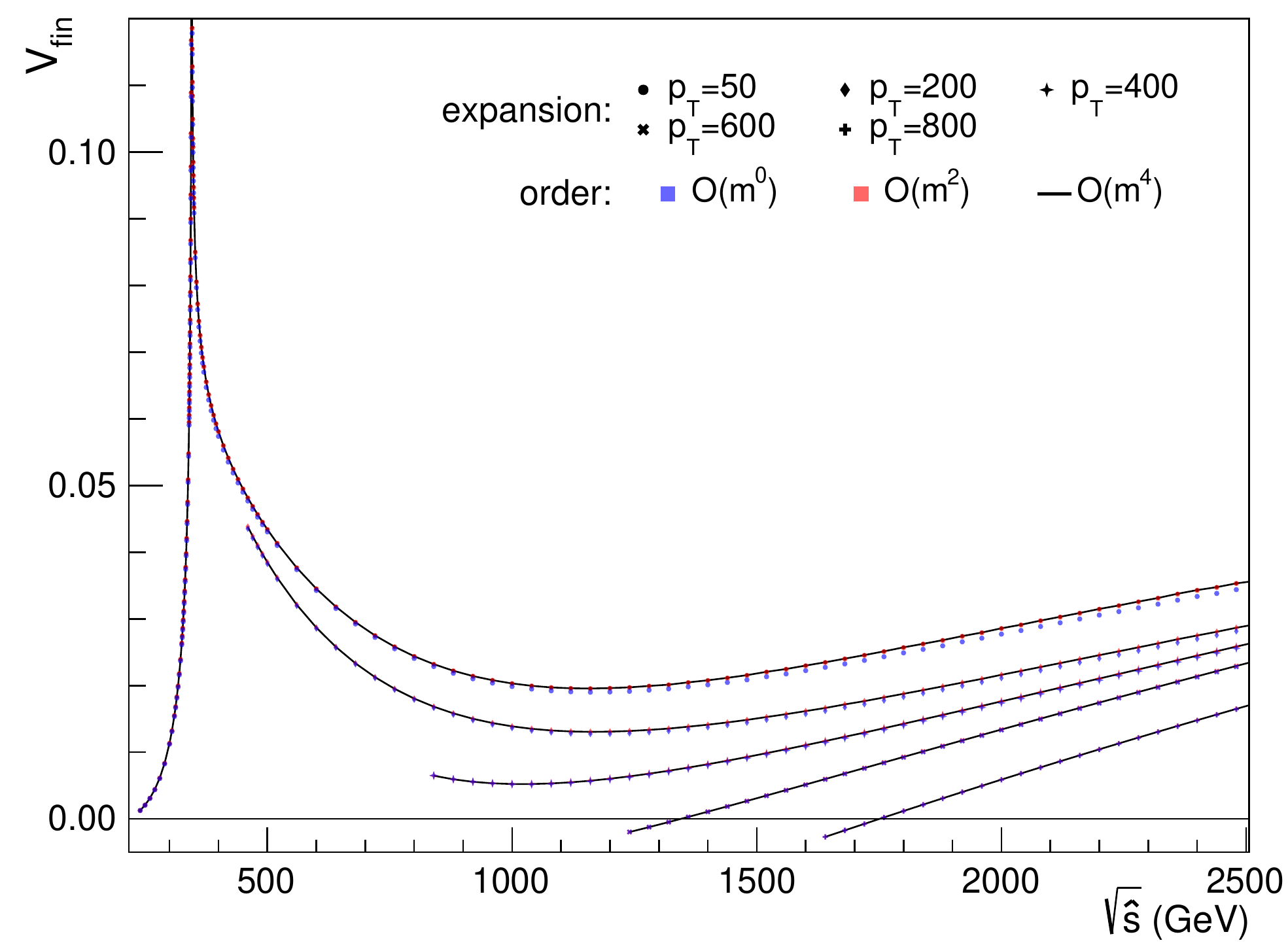}
\vspace{-2ex}
\caption{$\mathcal{V}_{\text{fin}}$ as a function of $\sqrt{\hat{s}}$ computed using the small-mass expansion up to $\mathcal{O}(m^0)$ (blue marks), $\mathcal{O}(m^2)$ (red marks), and $\mathcal{O}(m^4)$ (black lines) for several representative values of $p_T$.}
\label{fig:convergence}
\end{figure}

The validity of the small mass expansion has already been demonstrated in the case of Higgs boson pair production \cite{Wang:2020nnr}. To be more careful, we have performed a similar analysis for $ZH$ production. First of all, we have compared our results for the finite part of the virtual corrections with the results of Ref.~\cite{Chen:2020gae} obtained using the sector decomposition program \texttt{pySecDec}. It should be noted that the conventions for the finite part are slightly different between Ref.~\cite{Chen:2020gae} and our work, due to the different choices of the IR subtraction operator $\bm{I}(\epsilon)$\footnote{They have chosen $\tilde{K}_g = C_A\pi^2/2$.}. For comparison, we have performed the necessary transformation to arrive at their convention (dubbed $\mathcal{V}'_{\text{fin}}$ later). The results of $\mathcal{V}'_{\text{fin}}$ at six phase-space points are listed in Table~\ref{tab:amplitudeResults}. We find that our results at $\mathcal{O}(m^4)$ agree quite well with the \texttt{pySecDec} results. In particular, for the first four points where the \texttt{pySecDec} results are accurate enough, the relative errors of our results are much smaller than 0.1\%. We emphasize that our results can still be systematically improved by incorporating higher order terms in the small mass expansion.
The last two points correspond to the high energy region far above the $2m_t$ threshold, where the \texttt{pySecDec} results show large numeric uncertainties. On the other hand, the small mass expansion is expected to converge fast in the high energy region, and our results should be reliable here. The high energy expansion of Ref.~\cite{Davies:2020drs} also works in this region, and it will be interesting to compare our results with theirs as a crosscheck.

In Table~\ref{tab:amplitudeResults}, we also list the $\mathcal{O}(m^0)$ and $\mathcal{O}(m^2)$ results of our calculation, which shows the excellent convergence of the small mass expansion. To examine the convergence for a broader range of phase-space points, we show $\mathcal{V}_{\text{fin}}$ as a function of $\sqrt{\hat{s}}$ for several representative values of $p_T$ in Fig.~\ref{fig:convergence}. The blue and red marks represent the $\mathcal{O}(m^0)$ and $\mathcal{O}(m^2)$ results, respectively; while the black lines correspond to the $\mathcal{O}(m^4)$ ones. It can be seen that the $\mathcal{O}(m^2)$ and $\mathcal{O}(m^4)$ results almost overlap with each other completely, which demonstrates the reliability of the expansion in the entire phase-space. We expect that the terms at $\mathcal{O}(m^6)$ are irrelevant for phenomenological applications.

\begin{table*}[t!]
\centering
\footnotesize
\begin{tabular}{c||c||c|c|c||c|c}
$\mu_r = \mu_f$ & $\sigma^{gg}_{\rm LO}$ & $\sigma^{gg}_{\rm NLO}$ & $\sigma^{\text{w/o $gg$}}_{pp \to ZH}$ & $\sigma_{pp \to ZH}$ & $\sigma^{gg,m_t \to \infty}_{\rm NLO}$ & $\sigma_{pp \to ZH}^{m_t \to \infty}$ \\ \hline
$M_{ZH}/3$ & 73.56(7)    &  129.4(3) & 784.0(7) & 913.4(7) & 133.6(6) & 917.6(9) \\
$M_{ZH}$   & 51.03(5)    &  101.7(2) & 781.1(7) & 882.9(7) & 106.0(4) & 887.2(8) \\
$3M_{ZH}$  & 36.62(4)    &  \phantom{0}80.4(2) & 780.7(8) & 861.1(8) & \phantom{0}84.0(3) & 864.8(9)
\end{tabular}
\caption{The total cross sections (in fb) for $pp\to ZH$ and its subprocess $gg\to ZH$ at the 13~TeV LHC.
$\sigma^{\text{w/o $gg$}}_{pp \to ZH}$ is the cross section without the $gg\to ZH$ subprocess.
$\sigma^{gg}_{\rm LO}$ and $\sigma^{gg}_{\rm NLO}$ are the LO and NLO cross sections for $gg \to ZH$, in which the NLO contribution is one of the main new results of this work. $\sigma_{pp\to ZH} = \sigma^{\text{w/o $gg$}}_{pp \to ZH} + \sigma^{gg}_{\rm NLO}$ represents the state-of-the-art fixed-order predictions for this process. In the last two columns, we show for comparison the results in the heavy top limit.}
\label{tab:totalcrosssection3}
\end{table*}

We now combine the finite part of the virtual corrections with the IR-subtracted real corrections, and present our predictions for the total and differential cross sections. We first consider the LHC with a center-of-mass energy of $\sqrt{s} = 13$~TeV. We use the program package \texttt{vh@nnlo} \cite{Brein:2012ne, Harlander:2018yio} to calculate the contributions from the $q\bar{q}$ channel (including QCD and EW corrections). This program also gives the $gg \to ZH$ contributions up to the NLO in the heavy top limit, which we use as a reference to compare our results with. The various results for three values of $\mu_r = \mu_f$ are listed in Table~\ref{tab:totalcrosssection3}. As expected, the NLO corrections lead to significant enhancement (about 100\%) to the $gg\to ZH$ cross section. Combining our results with the $q\bar{q}$ contributions, we arrive at the state-of-the-art fixed-order prediction for the $pp \to ZH$ total cross section at the 13~TeV LHC:
\begin{equation}
\sigma_{pp \to ZH} = 882.9^{+3.5\%}_{-2.5\%}~\rm{fb} \, .
\end{equation}
In the last two columns of Table~\ref{tab:totalcrosssection3}, we show for comparison the results in the heavy top limit given by \texttt{vh@nnlo}. We find that the situation is quite different from the Higgs boson pair production: the finite top mass effects are much milder, which reduces the NLO cross sections in the $gg$ channel only by about 4\%. This accidental fact makes it possible that by calculating the $\mathcal{O}(\alpha_s^4)$ contributions in the heavy top limit, one could reduce the residue theoretical uncertainty of the total cross section down to the percent-level.

\begin{figure}[h!]
\centering
%\hspace{-4ex}
\includegraphics[width=0.6\linewidth]{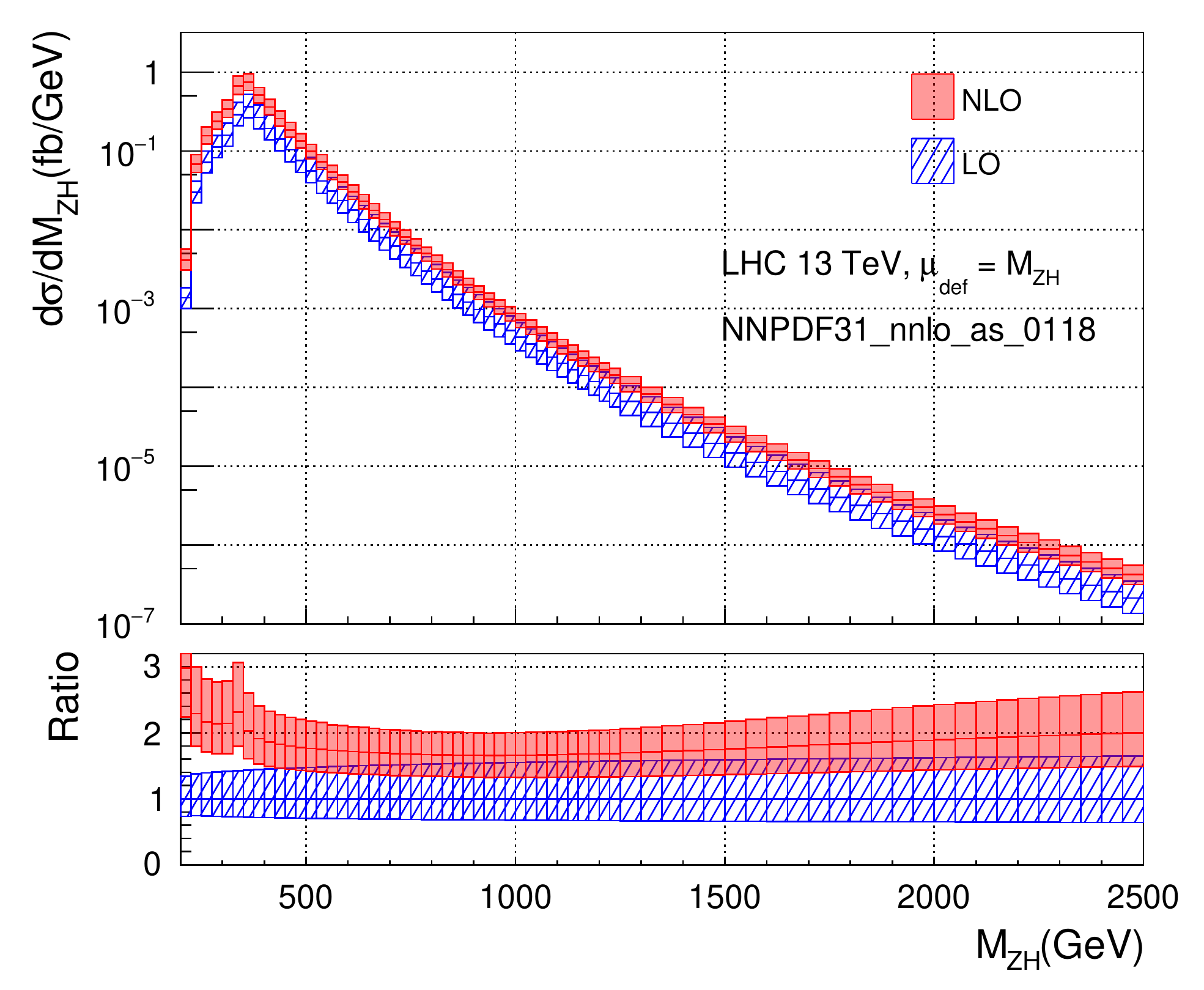}
\vspace{-2ex}
\caption{The LO and NLO differential cross sections in the $gg \to ZH$ channel with respect to the  $ZH$ invariant mass in the range $200~\mathrm{GeV} \leqslant M_{ZH}\leqslant 2500~\mathrm{GeV}$ at the 13 TeV LHC. The lower panel shows the ratios to the LO central values.}
\label{fig:mzh_distr_log}
\end{figure}

\begin{figure}[h!]
\centering
%\hspace{-4ex}
\includegraphics[width=0.6\linewidth]{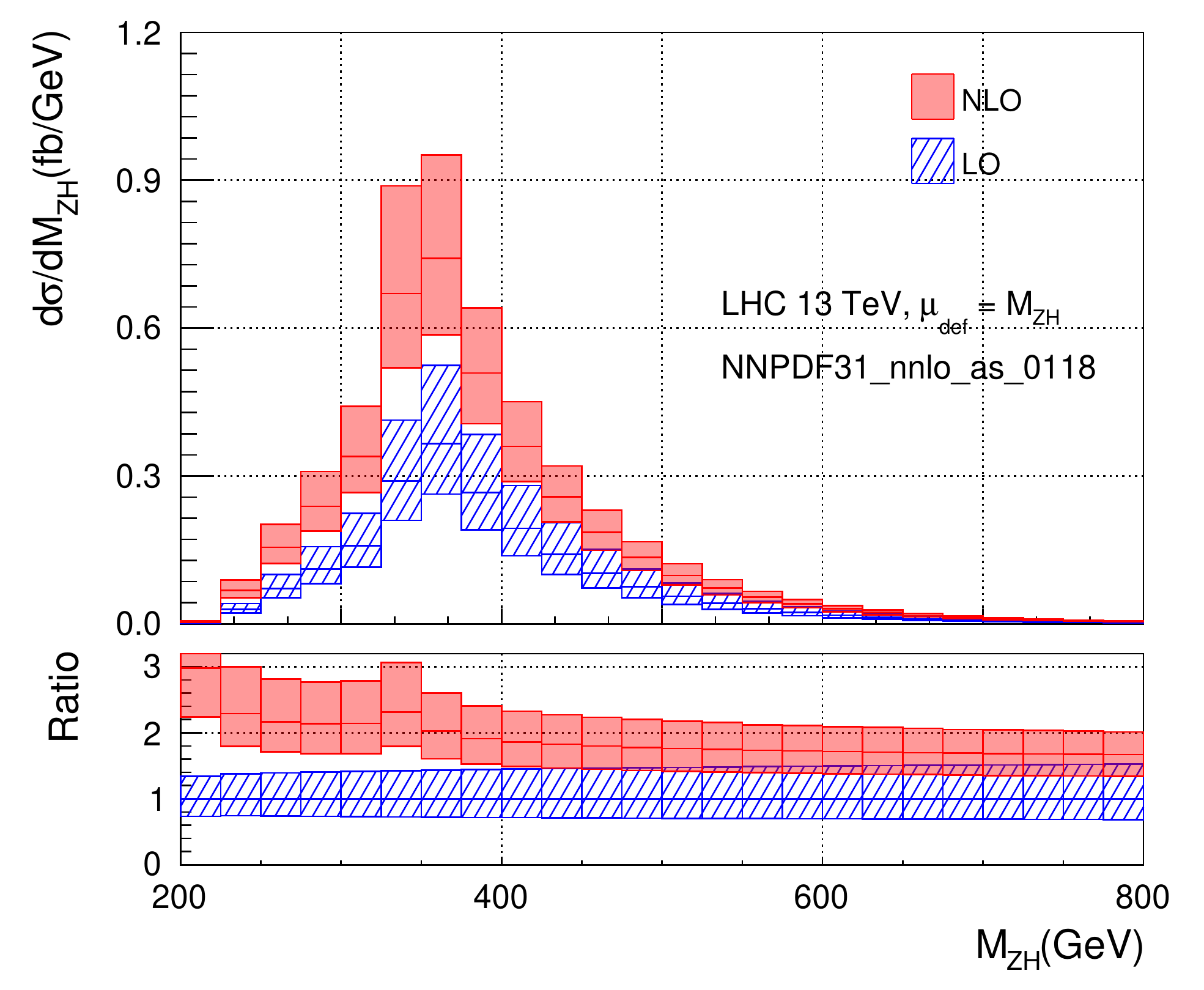}
\vspace{-2ex}
\caption{The LO and NLO differential cross sections in the $gg \to ZH$ channel with respect to the  $ZH$ invariant mass in the range  $200~\mathrm{GeV} \leqslant M_{ZH} \leqslant 800~\mathrm{GeV}$ at the 13~TeV LHC. The lower panel shows the ratios to the LO central values.}
\label{fig:mzh_distr_linear}
\end{figure}

We now turn to the differential cross sections. It is well-known that the heavy top limit is not valid above the $2m_t$ threshold. On the other hand, the small mass expansion provides reliable results for differential cross sections in the entire phase-space. As an example, we show in Fig.~\ref{fig:mzh_distr_log} the LO and NLO differential cross sections in the $gg \to ZH$ channel with respect to the invariant mass $M_{ZH}$ of the $Z$ boson and the Higgs boson at the 13~TeV LHC. The upper plot employs a logarithmic scale for the vertical axis to access the distributions in the broad range $200~\mathrm{GeV} \leqslant M_{ZH}\leqslant 2500~\mathrm{GeV}$, while the lower plot shows the ratios to the central values of the LO differential cross sections. It is clear that the sizes of the corrections are kinematics-dependent, and it is not sufficient to use a uniform $K$-factor to rescale the LO differential cross sections.
The NLO corrections are rather large across the whole range, especially around the peak and to the far tail. The significant corrections in the tail region have important implications for new physics searches, since new phenomena are usually most evident in the high energy regime. The boosted region is also relevant when using jet substructure techniques to measure the hadronic decays of the Higgs boson \cite{Butterworth:2008iy}.

The corrections in the peak region, on the other hand, are the most important to the total cross section. To see the peak region more clearly, we show the $M_{ZH}$ distributions in a narrower range in Fig.~\ref{fig:mzh_distr_linear}, with a linear vertical axis. It can be seen that the total cross section receives its most contributions from regions around the $2m_t$ threshold, where the NLO corrections are significant. The ratio plot also shows a small kink at the $2m_t$ threshold, which comes from the Coulomb-type enhancement in that region entering at NLO.

\begin{table*}[t!]
\centering
\footnotesize
\begin{tabular}{c||c||c|c|c||c|c}
$\mu_r = \mu_f$ & $\sigma^{gg}_{\rm LO}$ & $\sigma^{gg}_{\rm NLO}$ & $\sigma^{\text{w/o $gg$}}_{pp \to ZH}$ & $\sigma_{pp \to ZH}$ & $\sigma^{gg,m_t \to \infty}_{\rm NLO}$ & $\sigma_{pp \to ZH}^{m_t \to \infty}$ \\ \hline
$m_{ZH}/3$ & 310.5(3)    &  551(1) & 2106(2) & 2658(2) & 568(3) & 2675(3) \\
$m_{ZH}$   & 233.7(2)    &  451(1) & 2104(2) & 2555(2) & 476(2) & 2579(3) \\
$3m_{ZH}$  & 179.0(2)    &  373(1) & 2113(2) & 2485(2) & 396(2) & 2509(3)
\end{tabular}
\caption{Results for the 27~TeV HE-LHC similar to Table~\ref{tab:totalcrosssection3}.}
\label{tab:totalcrosssection27}
\end{table*}

Finally, we envision a possible future high-energy upgrade of the LHC (HE-LHC) operating at a center-of-mass energy of 27~TeV. In Table~\ref{tab:totalcrosssection27} we list the results for the total cross section at 27~TeV. Again, the NLO corrections are significant, with the top quark mass effect slightly larger than that in the 13~TeV case. The differential cross sections can also be easily computed, which we leave for future investigations.

\section{Conclusion}

In this work, we present for the first time a calculation of the complete NLO corrections to the $gg \to ZH$ process. We use the method of small mass expansion to tackle the most challenging two-loop virtual amplitude, in which the top quark mass dependence is retained throughout the calculations. We compare our results of the two-loop amplitude with the purely numeric results from \texttt{pySecDec}. We find that at phase-space points where \texttt{pySecDec} is precise enough, the relative deviations of our results are much smaller than $0.1\%$. We have also demonstrated the excellent convergence of the small mass expansion in the entire phase-space, which makes us confident that the expansion up to $\mathcal{O}(m^4)$ is sufficient for phenomenological applications.

We employ the dipole subtraction method to combine the virtual corrections with the real radiation contributions, and find that the IR divergences all cancel out. This allows us to give numeric predictions for the total and differential cross sections at the NLO. We add the contributions from the $q\bar{q}$ channel to obtain the state-of-the-art fixed-order predictions for the total cross sections, which amount to $\sigma_{pp \to ZH} = 882.9^{+3.5\%}_{-2.5\%}~\rm{fb}$ at the 13~TeV LHC, and $\sigma_{pp \to ZH} = 2.555^{+4.0\%}_{-2.7\%}~\rm{pb}$ at the 27~TeV HE-LHC.
We further present our results for a representative differential cross section: the invariant mass distribution of the $Z$ boson and the Higgs boson. We demonstrate that our method can provide reliable predictions for the differential cross sections from the low energy region all the way up to the highly-boosted regime. Our results are necessary ingredients towards reducing the theoretical uncertainties of the $pp \to ZH$ cross sections down to the percent-level, and provide important theoretical inputs for future precision experimental programs at the LHC and the HL-LHC. The results for other phenomenologically interesting distributions at the 13~TeV LHC, and the distributions at the 14~TeV LHC/HL-LHC and the 27~TeV HE-LHC will be presented in a forthcoming article.

\vspace{1ex}
This work was supported in part by the National Natural Science Foundation of China under Grant No. 11975030 and 11635001. The research of X. Xu was supported in part by the Swiss National Science Foundation (SNF) under Grant No. 200020\textit{$\_$}182038.
 
\bibliography{letter.bib}

\end{document}